\documentclass[apj]{emulateapj}

\begin{document}

\title{A direct empirical proof of the existence of dark matter
\footnote{Based on observations made with the NASA/ESA Hubble Space Telescope, 
obtained at the Space Telescope Science Institute, which is
operated by the Association of Universities for Research in Astronomy,
Inc., under NASA contract NAS 5--26555, under program 10200,
the 6.5 meter Magellan Telescopes located at Las Campanas Observatory, Chile,
the ESO Telescopes at the Paranal 
Observatories under program IDs 72.A-0511, 60.A-9203, and 64.O-0332,
and with the NASA Chandra X-ray 
Observatory, operated by the Smithsonian Astrophysics Observatory under
contract to NASA.}
}

\author{Douglas Clowe\altaffilmark{1}, Maru\v{s}a Brada\v{c}\altaffilmark{2},
Anthony H. Gonzalez\altaffilmark{3}, Maxim Markevitch\altaffilmark{4,}\altaffilmark{5}, Scott W. Randall\altaffilmark{4}, Christine Jones\altaffilmark{4}, and
Dennis Zaritsky\altaffilmark{1}}
\email{dclowe@as.arizona.edu}
\altaffiltext{1}{Steward Observatory, University of Arizona, 933 N Cherry Ave, 
Tucson, AZ 85721}
\altaffiltext{2}{Kavli Institute for Particle Astrophysics and Cosmology,
P.O.~Box 20450, MS29, Stanford, CA 94309}
\altaffiltext{3}{Department of Astronomy, University of Florida, 211 Bryant
Space Science Center, Gainesville, FL 32611}
\altaffiltext{4}{Harvard-Smithsonian Center for Astrophysics, 60 Garden St., Cambridge, MA 02138}
\altaffiltext{5}{Also Space Research Institute, Russian Acad.\ Sci.,
  Profsoyuznaya 84/32, Moscow 117997, Russia}

\submitted{ApJ Letters in press}

\begin{abstract}
We present new weak lensing observations of 1E0657$-$558 ($z=0.296$), a unique cluster 
merger, that enable a direct detection of dark matter, independent of 
assumptions regarding the nature of the gravitational force law.  Due to the 
collision of two clusters, the dissipationless stellar component and the 
fluid-like X-ray emitting plasma are spatially
segregated. By using both wide-field ground based images and HST/ACS images
of the cluster cores, we create gravitational lensing maps which show that 
the gravitational potential 
does not trace the plasma distribution, the dominant baryonic mass
component, but rather approximately traces the distribution of galaxies.  
An $8\sigma$ significance spatial offset of the center of the total mass 
from the center of the baryonic mass peaks cannot be explained
with an alteration of the gravitational force law, and thus proves that the 
majority of the matter in the system is unseen.
\end{abstract}

\section{Introduction}
We have known since 1937 that the gravitational potentials of  galaxy clusters 
are too deep to be caused by the detected baryonic mass and a Newtonian $r^{-2}$ 
gravitational force law \citep{ZW37.1}.  Proposed solutions either invoke 
dominant quantities of non-luminous ``dark matter''  \citep{OO32.1} or alterations to either 
the gravitational force law \citep{BE04.1,BR06.1} or the particles' dynamical 
response to it \citep{MI83.1}.  
Previous works aimed at distinguishing between the dark matter and
alternative gravity hypotheses in galaxies \citep{BU02.1,HO04.1} or galaxy 
clusters \citep{GA02.1,PO05.1} have used objects in which the
visible baryonic and hypothesized dark matter are spatially coincident, as
in most of the Universe.  These works favor the dark matter hypothesis, but
their conclusions were necessarily based on non-trivial assumptions such
as symmetry, the location of the center of mass of the system, and/or 
hydrostatic equilibrium, which left room for counterarguments.
The actual existence of dark 
matter can only be confirmed either by a laboratory detection or, in an 
astronomical context, by the discovery of a system in which the observed 
baryons and the inferred dark matter are spatially segregated.  An ongoing
galaxy cluster merger is such a system.  

\begin{figure*}
\plottwo{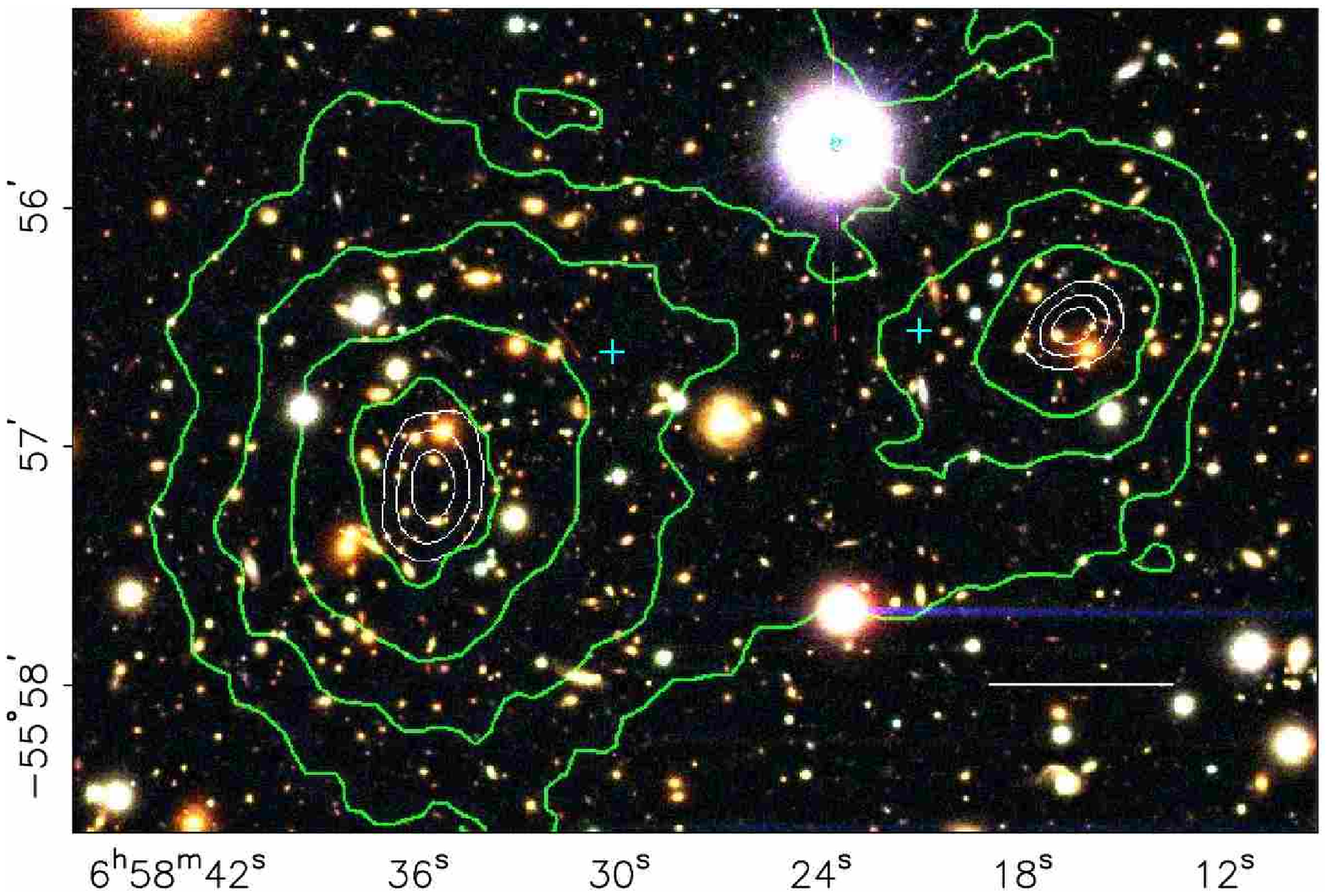}{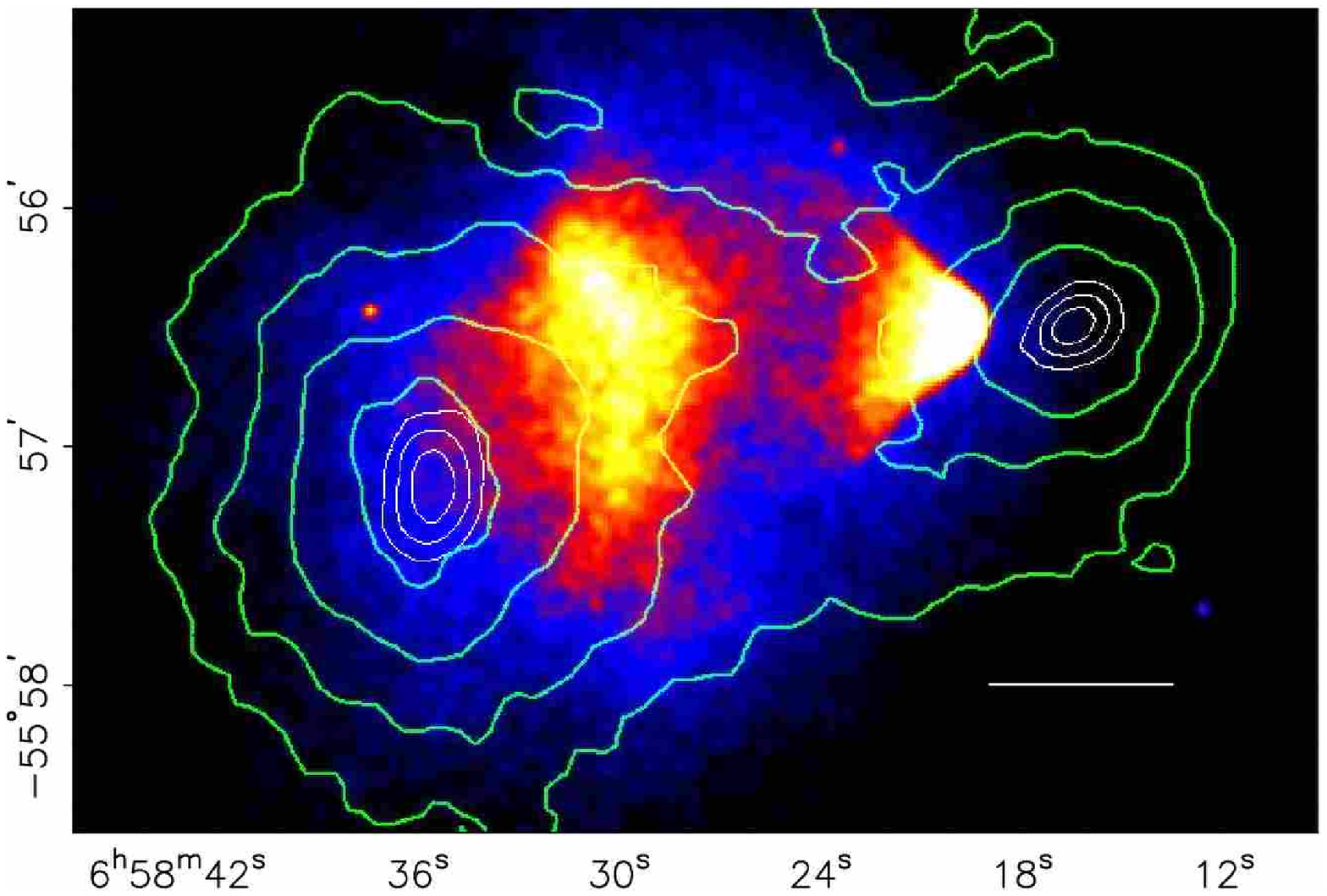}
\caption{
Shown above in the top panel is a color image from the Magellan images
of the merging cluster 1E0657$-$558, with the white bar indicating
200 kpc at the distance of the cluster.  In the bottom panel is a 500 ks 
Chandra image of the cluster.   Shown in green contours in both panels
are the weak lensing $\kappa$ reconstruction with the outer contour
level at $\kappa  =$ 0.16 and increasing in steps of 0.07.  The white
contours show the errors on the positions of the $\kappa$ peaks and correspond
to $68.3\%$, $95.5\%$, and $99.7\%$ confidence levels.  The blue $+$s show
the location of the centers used to measure the masses of the plasma
clouds in Table \ref{table2}.}
\label{fig1}
\end{figure*}

Given sufficient time, galaxies (whose stellar component makes up 
$\sim 1-2 \%$ of the mass \citep{KO03.1} under the assumption of Newtonian 
gravity), plasma ($\sim 5-15$\% of the mass \citep{AL02.1,VI06.1}), 
and any dark matter in a typical cluster 
acquire similar, centrally-symmetric spatial distributions tracing the common 
gravitational potential.  However, during a merger of two clusters, galaxies 
behave as collisionless particles, while the fluid-like X-ray emitting 
intracluster 
plasma experiences ram pressure.  Therefore, in the course of a cluster 
collision, galaxies spatially decouple from the plasma.  We clearly see this 
effect in the unique cluster 1E0657$-$558 \citep{TU98.1}.

The cluster has two primary galaxy concentrations separated by 0.72 Mpc 
on the sky, 
a less massive ($T\sim 6$ keV) western subcluster and a more massive 
($T\sim 14$ keV) eastern main cluster \citep{MA02.1}.   Both concentrations 
have associated X-ray emitting plasma offset from the galaxies 
toward the center of the system.  
The X-ray image also shows a prominent bow shock on the western side
of the western plasma cloud, indicating that the subcluster is currently
moving away from the main cluster at $\sim$4700 km/s.  As the line-of-sight
velocity difference between the components is only 
$\sim 600 $ km/s \citep{BA02.1},
the merger must be occurring nearly in the plane of the sky and the
cores passed through each other $\sim 100$ Myr ago. 

Two galaxy concentrations that correspond to the main cluster and the
smaller subcluster have moved ahead of their respective plasma clouds 
that have been slowed by ram pressure.  This phenomenon provides an excellent 
setup for our simple test.  In the absence of dark matter, the gravitational 
potential will trace 
the dominant visible matter component, which is the X-ray plasma.  If, on the 
other hand, the mass is indeed dominated by collisionless dark matter, the 
potential will trace the distribution of that component, which is
expected to be spatially coincident with the collisionless galaxies.  Thus,
by deriving a map of the gravitational potential, one can discriminate
between these possibilities.  We published an initial attempt at this
using an archival VLT image \citep{CL04.1}; here we add three additional
optical image sets which allows us to increase the significance of the
weak lensing results by more than a factor of 3.

In this paper, we measure distances at the redshift of the cluster, $z=0.296$,
by assuming 
an $\Omega_\mathrm{m} =0.3, \lambda =0.7, H_0=70\mathrm{km/s/Mpc}$ cosmology 
which results in $4.413$ kpc/$\arcsec$ plate-scale.  None of the results of
this paper are dependent on this assumption; changing the assumed cosmology
will result in a change of the distances and absolute masses measured,
but the relative masses of the various structures in each measurement
remain unchanged.

\section{Methodology and Data}

\begin{deluxetable*}{llcccccc}
  \tabletypesize{\small}
  \tablecaption{Optical Imaging Sets\label{table1}}
\tablehead{\colhead{Instrument} & \colhead{Date of Obs.} & \colhead{FoV} & \colhead{Passband}  & 
\colhead{$t_{\mathrm{exp}}$ (s)}& \colhead{$m_{\mathrm{lim}}$} & 
\colhead{$n_\mathrm{d}$ ($\arcmin^{-2}$)} & \colhead{seeing}\\}
\startdata
2.2m ESO/MPG & $01/2004$ & $34\arcmin \times 34\arcmin$ & R  &  14100 & $23.9$ & 15 & $0\farcs8$ \\
Wide Field Imager & $01/2004$ & & B & 6580 & & & $1\farcs0$ \\
 & $01/2004$ & & V & 5640 & & & $0\farcs9$ \\
6.5m Magellan & $01/15/2004$ & $8\arcmin$ radius & R & 10800 & $25.1$ & 35 & $0\farcs6$\\
IMACS & $01/15/2004$ & & B & 2700 & & & $0\farcs9$\\
 & $01/15/2004$ & & V & 2400 & & & $0\farcs8$\\
HST ACS & $10/21/2004$ & 3\farcm 5$\times $3\farcm 5 & F814W & 4944 & $27.6$ & 87 & $0\farcs12$ \\
subcluster & $10/21/2004$ & & F435W & 2420 & & & $0\farcs12$ \\
 & $10/21/2004$ & & F606W & 2336 & & & $0\farcs12$ \\
main cluster & $10/21/2004$ & 3\farcm 5$\times $3\farcm 5 & F606W & 2336 & $26.1$ & 54 & $0\farcs12$ \\
\enddata
\tablecomments{Limiting magnitudes for completion are given for galaxies and measured by 
where the number counts depart from a power law.  All image sets had objects
detected in the reddest passband available.}
\end{deluxetable*}

We construct a map of the gravitational potential using weak gravitational
lensing \citep{ME99.1,BA01.1}, which measures the distortions of images
of background galaxies caused by the gravitational deflection of light
by the cluster's mass.  This deflection
stretches the image of the galaxy preferentially in the direction
perpendicular to that of the cluster's center of mass.
The imparted ellipticity is typically comparable to or smaller than that 
intrinsic to
the galaxy, and thus the distortion is only measurable statistically
with large numbers of background galaxies.  To do this measurement, we detect faint 
galaxies on deep optical images and calculate an ellipticity from the
second moment of their surface brightness distribution, correcting
the ellipticity for smearing by the point spread function
(corrections for both anisotropies and smearing are obtained
using an implementation of the KSB technique \citep{KA95.1} discussed in 
\citet{CL06.1}).  The corrected ellipticities are a direct, but
noisy, measurement of the reduced shear $\vec{g} = \vec{\gamma}/(1-\kappa)$.  
The shear $\vec{\gamma}$ is the amount of anisotropic stretching of the
galaxy image.  The convergence $\kappa$ is the shape-independent
increase in the size of the galaxy image.  In Newtonian gravity, $\kappa$ is equal 
to the surface mass density of the lens divided by a scaling constant. 
In non-standard gravity models, $\kappa$ is no longer linearly related 
to the surface density but is instead a non-local function that 
scales as the mass raised to a power less than one for a planar lens, reaching 
the limit of one half for constant 
acceleration \citep{MO01.1,ZH06.1}.  While one can no longer directly obtain a 
map of the surface mass density using the distribution of $\kappa$ 
in non-standard gravity models, the locations of the $\kappa$ peaks,
after adjusting for the extended wings,
correspond to the locations of the surface mass density peaks.

Our goal is thus to obtain a map of $\kappa$.
One can combine derivatives of $\vec{g}$ to obtain \citep{SC95.1,KA95.2} 
\[
\nabla \ln (1 - \kappa) = {1\over 1 - g_1^2 - g_2^2} \left( {\begin{array}{cc}
1 + g_1 & g_2 \\ g_2 & 1 - g_1 \end{array}} \right) \left ( {\begin{array}{c}
g_{1,1} + g_{2,2} \\ g_{2,1} - g_{1,2} \end{array}} \right),
\]
which is integrated over the data field and converted into a two-dimensional
map of $\kappa$.  The observationally unconstrained constant of integration,
typically referred to as the ``mass-sheet degeneracy,'' is effectively 
the true mean of $\ln(1-\kappa)$ at the edge of the reconstruction.  This method does,
however, systematically underestimate $\kappa$ in the cores of massive 
clusters.  This results in a slight 
increase to the centroiding errors of the peaks, and our measurements of 
$\kappa$ in the peaks of the components are only lower bounds.

For 1E0657$-$558, we have accumulated an exceptionally rich optical dataset,
which we will use here to measure $\vec{g}$.  It consists of the four sets of 
optical images shown in Table~\ref{table1} and the VLT image set used 
in \citet{CL04.1}; 
the additional images significantly increase the
maximum resolution obtainable in the $\kappa$ reconstructions due to the 
increased number of background galaxies, particularly in the area
covered by the ACS images, with which we measure the reduced shear.  We reduce 
each image set independently and create galaxy catalogs with 3 passband 
photometry.  The one exception is the single passband HST pointing of
main cluster, for which we measure colors from the Magellan images.
Because it is not feasible to measure redshifts for all galaxies in
the field, we select likely background galaxies using magnitude and color cuts
($m_{814} > 22$ and not in the rhombus defined by $0.5 < m_{606}-m_{814} < 
1.5$, $m_{435}-m_{606} > 1.5 * (m_{606}-m_{814}) - 0.25$, and 
$m_{435}-m_{606} < 1.6 * (m_{606}-m_{814}) + 0.4$ for the ACS images; similar
for the other image sets)
that were calibrated with photometric redshifts from the HDF-S \citep{FO99.1}.
Each galaxy has a statistical weight based on its significance of detection 
in the image set \citep{CL06.1}, and the weights
are normalized among catalogs by comparing the rms reduced shear measured
in a region away from the cores of the cluster common to all five data sets.
To combine the catalogs, we adopt a weighted average of the reduced
shear measurements and appropriately increase the statistical weight 
of galaxies that occur in more than one catalog.  

\section{Analysis}

We use the combined catalog to create a two-dimensional $\kappa$
reconstruction, the central portion of which is shown in Fig.~\ref{fig1}.  
Two major peaks are clearly visible in the reconstruction, one centered 
$7\farcs1$ east and $6\farcs5$ north of the subcluster's 
brightest cluster galaxy (BCG) and 
detected at $8\sigma$ significance (as compared to $3\sigma$ in 
\citep{CL04.1}), and one centered $2\farcs5$ east and
$11\farcs5$ south of the northern BCG in the main cluster
($21\farcs2$ west and $17\farcs7$ north of the southern BCG) 
detected at $12\sigma$.  We estimate centroid uncertainties by repeating
bootstrap samplings of the background galaxy catalog, performing
a $\kappa$ reconstruction with the resampled catalogs, and measuring the
centroid of each peak.  Both peaks are offset from their respective BCG 
by $\sim 2\sigma$, but are within $1\sigma$ of the luminosity centroid
of the respective component's galaxies (both BCGs are slightly offset
from the center of galaxy concentrations).  Both peaks are also offset at 
$\sim 8\sigma$
from the center of mass of their respective plasma clouds.  They are
skewed toward the plasma clouds, which is expected because
the plasma contributes about 1/10th of the total cluster mass \citep{AL02.1,VI06.1}
(and a higher fraction in non-standard gravity models without dark matter).  
The skew in each $\kappa$ peak
toward the X-ray plasma is significant even after correcting for the
overlapping wings of the other peak, and the degree of skewness is
consistent with the X-ray plasma contributing $14\%^{+16\%}_{-14\%}$ of the observed
$\kappa$ in the main cluster and $10\%^{+30\%}_{-10\%}$ in the subcluster
(see Table~\ref{table2}).  
Because of the large size of the reconstruction ($34\arcmin$ or
9 Mpc on a side), the change in $\kappa$ due to the mass-sheet
degeneracy should be less than $1\%$ and any
systematic effects on the centroid and skewness of the peaks are much 
smaller than the measured error bars.  

\begin{deluxetable*}{lccccc}
\tabletypesize{\small}
\tablecaption{Component Masses\label{table2}}
\tablehead{\colhead{Component} & RA (J2000) & Dec (J2000) & \colhead{$M_{\mathrm{X}} (10^{12} M_\odot)$} &
\colhead{$M_{\ast} (10^{12} M_\odot)$} & \colhead{$\bar{\kappa}$} \\}
\startdata
Main cluster BCG & $06:58:35.3$ & $-55:56:56.3$ & $5.5\pm 0.6$ & $0.54\pm 0.08$ & $0.36\pm 0.06$\\
Main cluster plasma & $06:58:30.2$ & $-55:56:35.9$ & $6.6\pm 0.7$ & $0.23\pm 0.02$ & $0.05\pm 0.06$\\
Subcluster BCG & $06:58:16.0$ & $-55:56:35.1$ & $2.7\pm 0.3$ & $0.58\pm 0.09$ & $0.20\pm 0.05$\\
Subcluster plasma & $06:58:21.2$ & $-55:56:30.0$ & $5.8\pm 0.6$ & $0.12\pm 0.01$ & $0.02\pm 0.06$\\ 
\enddata
\tablecomments{All values are calculated by averaging over an aperture of 100 
kpc radius around the given position (marked with blue $+$s for the
centers of the plasma clouds in Fig~\ref{fig1}).  $\bar{\kappa}$ measurements
for the plasma clouds are the residual left over after subtraction of
circularly symmetric profiles centered on the BCGs.
}
\end{deluxetable*}

The projected cluster galaxy stellar mass and plasma mass within 100 kpc 
apertures centered on the BCGs and X-ray plasma peaks are shown in 
Table~\ref{table2}.  This aperture size was chosen as smaller
apertures had significantly higher kappa measurement errors and
larger apertures resulted in significant overlap of the apertures.
Plasma masses were computed from a multicomponent
3-dimensional cluster model fit to the Chandra X-ray image
(details of this fit will be given elsewhere). The emission
in the Chandra energy band (mostly optically-thin thermal
bremsstrahlung) is proportional to the square of the plasma
density, with a small correction for the plasma temperature
(also measured from the X-ray spectra), which gives the
plasma mass. Because of the simplicity of this cluster's
geometry, especially at the location of the
subcluster, this mass estimate is quite robust (to a 10\%
accuracy).

Stellar masses are calculated from the $I$-band luminosity of all galaxies 
equal in brightness or fainter than the component BCG.  The luminosities 
were converted into mass assuming \citep{KA03.1} $M/L_I = 2$.  The
assumed mass-to-light ratio is highly uncertain (can vary between 0.5
and 3) and depends on the history of recent star formation of the galaxies 
in the apertures; however even in the case of an extreme deviation, the
X-ray plasma is still the dominant baryonic component in all of the apertures.
The quoted errors are only the errors on measuring the luminosity and do not 
include the uncertainty in the assumed mass-to-light ratio.  Because we
did not apply a color selection to the galaxies, these measurements
are an upper limit on the stellar mass as they include contributions
from galaxies not affiliated with the cluster.

The mean $\kappa$ at each BCG was calculated by fitting a two 
peak model, each peak circularly symmetric, to the reconstruction and
subtracting the contribution of the other peak at that distance.
The mean $\kappa$ for each plasma cloud is the excess $\kappa$ after
subtracting off the values for both peaks.

The total of the two visible mass components of the subcluster
is greater by a factor of 2 at the plasma peak than at the BCG; 
however, the center of the lensing mass is located near the BCG. The
difference of the baryonic mass between these two positions
would be even greater if we excluded a contribution of the
non-peaked plasma component between the shock front and the
subcluster. For the main cluster, we see the same effect,
although the baryonic mass difference is smaller. Note that
both the plasma mass and the stellar mass are determined
directly from the X-ray and optical images, respectively,
independently of any gravity or dark matter models.

\section{Discussion}

A key limitation of the gravitational lensing methodology is that it only
produces a two-dimensional map of $\kappa$, and hence raises the possibility
that structures seen in the map are caused by physically unrelated masses
along the line-of-sight.  Because the background galaxies reside at a 
mean $z\sim1$, structures capable of providing a significance amount of 
$\kappa$ must lie at 
$z\la 0.8$.  By comparing the measured shear for galaxies divided 
into crude redshift bins using photometric redshifts \citep{WI03.1}, we 
further limit the 
redshift of the lensing objects to $0.18<z<0.39$. This range is consistent
with the cluster redshift, but corresponds to a large volume
in which a structure unassociated with the cluster could exist and be projected
onto the lensing map.  However, the number density of structures with 
these lensing strengths in blank field surveys is $\sim 10^{-3}$ 
arcmin$^{-2}$ for the subcluster \citep{WI06.1}, and an order of 
magnitude less for the main cluster, resulting in a $\sim 10^{-7}$ probability 
of having two structures within a square arcminute 
of the observed cluster cores.  Further, all such lenses observed in 
these cosmic shear surveys
are clusters with enough plasma and galaxies to be easily observable.  
There is no evidence, however, in our deep imaging for additional cluster
sized concentrations of galaxies or of plasma hotter than
$T\sim 0.5$ keV (the lower bound of the Chandra energy band) near the
observed lensing structures.

Another alternate explanation of the lensing signal is related to the
fact that clusters form at the intersections of matter filaments \citep{BO96.1}.
In principle, one could imagine two line-of-sight filaments of
intergalactic gas (too cool to be visible with Chandra and
too diffuse to have cooled into stars) extending from the cluster
at the locations of the weak lensing peaks. To explain the measured surface 
mass density, such filaments would have to be several Megaparsecs
long, very narrow, and oriented exactly along the line of
sight.  The probability of such an orientation for two such filaments
in the field is $\sim 10^{-6}$.
Further, because the two cluster components are moving
at a relative transverse velocity of 4700 km/s compared to the typical
peculiar velocities in the CMB frame of a few hundred km/s, the filaments
could coincide so exactly with each of the BCGs only by chance.  This is
an additional factor of $\sim 10^{-5}$ reduction in probability.
While such projections become
more important in non-standard gravity models because in such models the
thin lens approximation breaks down \citep{MO01.1} and structures with a given
surface density produce a greater amount of lensing the more they are
extended along the line-of-sight, two such projections would still have a
$\ll 10^{-8}$ probability.  
Finally, we mention that two other merging clusters, MS1054$-$03 \citep{JE05.1}
and A520 (in preparation), exhibit similar offsets between the
peaks of the lensing and baryonic mass, although based on lensing
reconstructions with lower spatial resolution and less clear-cut cluster
geometry.

A final possibility is that some alternative gravity models may be able
to suppress the lensing potential of the central peak in a multiple-peak
system, as in \citet{AN06.1}.  That work used a model of a gas disk
located between two symmetric mass concentrations representing the
galaxy subclusters.  In their $\kappa$ map, derived in the TeVeS
framework \citep{BE04.1}, the relative signal from this disk may
be suppressed, but would still be easily visible with the noise
levels of our reconstruction.  Our $\kappa$ map, however, has
no evidence of any mass concentration between the two galaxy
subclusters other than the small perturbations consistent with
the gas mass contribution in Newtonian gravity.  Furthermore, such
a suppression has also only been shown to work for symmetric systems which 
have the central peak directly between the two outer peaks.
In 1E0657$-$558, however, the
X-ray plasma, which would provide the central peak, lies north
of the line connecting the two $\kappa$ peaks.  Further, the absolute
$\kappa$ levels of the peaks observed in 1E0657$-$558 are in good
agreement with those in systems with similar velocity dispersions 
and X-ray temperatures \citep[e.g.][]{CL02.1} which have the gas
and the galaxies coincident.  The $\kappa$-to-light ratios are also
consistent with those in normal clusters with coincident gas and galaxies.
Therefore one would need to not only suppress the inner peak in
the $\kappa$ map relative to the two outer peaks in this system,
but also enhance the strength of 
the outer peaks to make up for the missing plasma mass.

Any non-standard gravitational force that scales with baryonic mass will
fail to reproduce these observations.  The lensing peaks require unseen
matter concentrations that are more massive than and offset from the plasma. 
While the existence of dark matter removes the primary motivation for 
alternative gravity models, it does not preclude non-standard gravity.
The scaling relation between $\kappa$ and surface mass density, however, 
has important consequences for models that mix dark matter with non-Newtonian 
gravity:  to achieve the $\sim 7:1$ ratio in $\kappa$ between the dark matter 
+ galaxy component and the plasma component (Table~\ref{table2}), the true 
ratio of mass would be even higher (as high as 49:1 for a constant 
acceleration model; although MOND \citep{MI83.1} would not reach this ratio 
as the dark matter density would become high enough to shift the acceleration 
into the quasi-Newtonian regime), making the 
need for dark matter even more acute.  Such high concentrations of dark matter,
however, are extremely unlikely based on the measured X-ray plasma 
temperatures \citep{MA02.1} and cluster galaxy velocity dispersions \citep{BA02.1}.

The spatial separation of the dominant baryonic component in a galaxy cluster 
from the hypothesized dark matter produced during a cluster merger has enabled 
us to directly compare the dark matter hypothesis to one with only visible
matter but a modified law of gravity.  The observed displacement 
between the bulk of the baryons and the gravitational potential proves the 
presence of dark matter for the most general assumptions
regarding the behavior of gravity.

\acknowledgements
We wish to thank Bhuvnesh Jain, Hongsheng Zhao, and Peter Schneider for
useful discussions.  This work was supported by NASA through grant number
LTSA04-0000-0041 (DC and DZ).
Support for program 10200 was provided by NASA through grant GO-10200.01 from 
the Space Telescope Science Institute, which is operated by the Association of 
Universities for Research in Astronomy, Inc., under NASA contract NAS 5-26555.
MB acknowledges support from the NSF grant AST-0206286.  MM acknowledges
support from Chandra grant GO4-5152X.

\end{document}